\documentclass[twocolumn,prd,superscriptaddress,showpacs,preprintnumbers,amsmath,amssymb,nofootinbib]{revtex4}

\usepackage{exscale}                  
\usepackage{amsfonts}
\usepackage{amssymb,amscd}
\usepackage[dvips]{epsfig}                   
\usepackage{array}
\usepackage{wrapfig}
\usepackage{dsfont}
\usepackage{color}
\newcommand{\beqa}{\begin{eqnarray}}
\newcommand{\eeqa}{\end{eqnarray}}
\newcommand{\beq}{\begin{equation}}
\newcommand{\eeq}{\end{equation}}
\newcommand{\bk}{\bar{\kappa}}
\newcommand{\dk}{\Delta\kappa}

\def\di{\displaystyle}

\def\bg{\begin{eqnarray}\begin{array}{rcl}\displaystyle}
\def\eg{\end{array} &\di    &\di   \end{eqnarray}}
\def\bm#1{\begin{eqnarray}\begin{array}{#1}\di}
\def\bmo#1{\begin{eqnarray*}\begin{array}{#1}\di}
\def\bml#1#2{\begin{eqnarray}\begin{array}{#1}\label{#2}\di}
\def\bgo{\begin{eqnarray*}\begin{array}{rcl}\displaystyle}
\def\ego{\end{array} &\di    &\di \nonumber  \end{eqnarray*}}

\def\btensor#1#2{\renew\left#1\begin{array}{#2}\di}
\def\brtensor#1#2#3{\ren#3\left#1\begin{array}{#2}}
\def\botensor#1#2{\renew\left#1\begin{array}{#2}}
\def\etensor#1{\end{array}\right#1}

\def\eq#1{(\ref{#1})}
\def\Eq#1{Eq.~(\ref{#1})}

\def\Tr{{\rm Tr}}

\def\s0#1#2{\mbox{\small{$ \frac{#1}{#2} $}}}
\def\0#1#2{\frac{#1}{#2}}

\def\N{{{\rm l}\!{\rm N}}}

\def\Z{\mathds{Z}}
\def\N{\mathds{N}}

\definecolor{christian}{rgb}{0,0.4,0.7}
\definecolor{jan}{rgb}{0.7,0,0.4}
\definecolor{comment}{rgb}{0.9,0,0}

\begin{document}
\title{Uniqueness of infrared asymptotics in Landau gauge Yang-Mills theory II} 
\author{Christian~S.~Fischer}
\affiliation{Institut f\"ur Kernphysik, 
  Technische Universit\"at Darmstadt,
  Schlossgartenstra{\ss}e 9,\\ 
  D-64289 Darmstadt, Germany}
\affiliation{GSI Helmholtzzentrum f\"ur Schwerionenforschung GmbH, 
  Planckstr. 1  D-64291 Darmstadt, Germany.}
\author{Jan M.~Pawlowski}
\affiliation{Institut f\"ur Theoretische Physik, University of
  Heidelberg, Philosophenweg 16, D-62910 Heidelberg, Germany.}
\affiliation{ExtreMe Matter Institute EMMI, GSI Helmholtzzentrum f\"ur Schwerionenforschung mbH, 
  Planckstr. 1  D-64291 Darmstadt, Germany.}
\begin{abstract}
  We present a shortened and simplified version of our proof
  \cite{Fischer:2006vf} of the uniqueness of the scaling solution for
  the infrared asymptotics of Green functions in Landau gauge
  Yang-Mills theory. The simplification relates to a new RG-invariant
  arrangement of Green functions applicable to general theories. As
  before the proof relies on the necessary consistency between
  Dyson-Schwinger equations (DSEs) and functional renormalisation
  group equations (FRGs). We also demonstrate the existence
  of a specific scaling solution for both, DSEs and FRGs, that
  displays uniform and soft kinematic singularities.
\end{abstract}


\maketitle

\section{Introduction \label{Intro}}

The infrared behaviour of strongly interacting quantum field theories
is of general interest. In particular the infrared behaviour of Landau
gauge Yang-Mills theory is a much debated issue in the past years.
This is due to the close connection of the ghost and gluon propagators
to confinement scenarios proposed by Kugo and Ojima \cite{Kugo},
Gribov \cite{Gribov} and Zwanziger \cite{gzwanziger2}.  Within
functional methods a self-consistent infrared asymptotic solution of
the whole tower of Dyson-Schwinger equations (DSEs) and functional
renormalisation group equations (FRGs) has been found \cite{von
  Smekal:1997vx,Lerche:2002ep,Pawlowski:2003hq,Alkofer:2004it,%
  Fischer:2006vf,Huber:2007kc} that supports these scenarios and is
also consistent with global BRST symmetry \cite{Fischer:2008uz}. This
{\it scaling} solution implies that all Greens functions scale like a
power of momentum in the infrared with interrelated anomalous
dimensions if all momenta are scaled. That entails the absence of mass
scales below which some degrees of freedom decouple.  In such a
scenario the scaling power of Green functions can be extracted by a
power counting analysis. As a further direct consequence all couplings
have fixed points at zero momentum.

An alternative infrared solution of Yang-Mills theory is provided by
the {\it decoupling} solution 
\cite{Cornwall:1981zr,Aguilar:2008xm,Boucaud:2008ji,Dudal:2008sp}.
This type of
solutions has been discussed in detail in \cite{Fischer:2008uz}, and
has been shown to be inconsistent with global BRST symmetry. In the
present context it can be interpreted as the decoupling of (part of)
the propagating degrees of freedom below a mass scale. In such a case
the infrared asymptotics cannot be fixed uniquely by a scaling analysis.

In \cite{Fischer:2006vf} we suggested a combined analysis of the towers 
of DSEs and FRGs for an infrared scaling analysis, being applicable to
general theories. We have shown that apart from decoupling there is
only one, unique scaling solution of infrared Yang-Mills theory in
Landau gauge.  Here we present a greatly simplified version of our
proof which also allows the reduction of the number of
presuppositions. The result, of course, is the same as in
\cite{Alkofer:2004it,Fischer:2006vf,Huber:2007kc}.  In addition we
present an explicit scaling solution for DSEs and FRGs that involves
also kinematical singularities \cite{Alkofer:2008jy}. The knowledge of
these kinematical singularities is not necessary for the proof of the
existence and uniqueness of the global scaling, and were not discussed
explicitly in our previous work \cite{Fischer:2006vf}.

In Section~\ref{sec:fun} we introduce FRG and DSE equations for the
effective action and discuss momentum and RG scaling for the effective
action and its vertices. With help of Appendix~\ref{app:fun&scale} the
functional equations are written in a similar form. A convenient
parameterisation of the vertices is introduced that splits off the
renormalisation group (RG)-scaling and reduces the discussion to that
of the scaling properties of RG-invariant quantities. This very
natural reduction is the key ingredient of the simplification of the
proof, and is applicable to general theories. It also allows us to provide
heuristic arguments for the existence of a unique scaling solution,
that should facilitate the following of the proof. In 
Section~\ref{sec:proof} we derive the respective scaling constraints
from FRG and DSE equations, the combination of which provides a unique
scaling solution. In Section~\ref{sec:kinematics} we extend our
analysis to kinematical singularities of vertices, the details can be
found in Appendix~\ref{app:vertices}. We close with a short summary of
our findings.

\section{Functional relations for the effective action}\label{sec:fun} 

\begin{figure}[h]
\centerline{\includegraphics[width=\columnwidth]{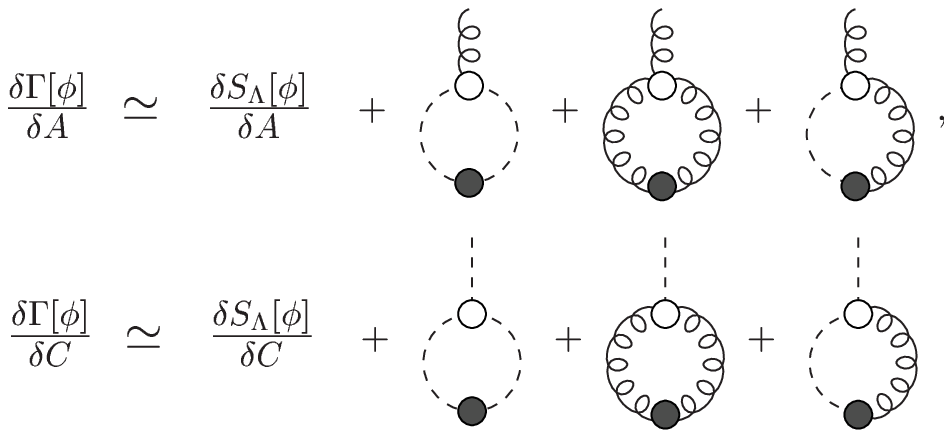}}
\caption{Infrared asymptotics of the FRG. Filled circles denote fully
  dressed field dependent propagators.  Empty circles denote fully
  dressed field dependent vertices.}
\label{fig:funFRG}
\end{figure}

\begin{figure}[h]
\centerline{\includegraphics[width=\columnwidth]{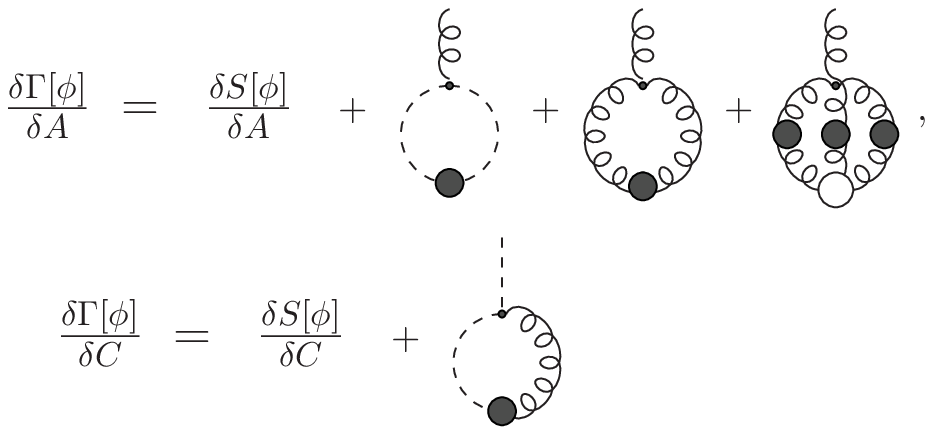}}
\caption{Functional Dyson-Schwinger equation (DSE) for the effective 
action. Filled circles denote fully dressed field dependent propagators. 
Empty circles denote fully dressed field dependent vertices, 
dots denote field dependent bare vertices.}
\label{fig:funDSE}
\end{figure}

The starting point of our analysis is the functional form of FRGs and
DSEs, depicted in Fig.~\ref{fig:funFRG} and Fig.~\ref{fig:funDSE}
respectively.
In Fig.~\ref{fig:funFRG} we have rewritten the FRG in a form similar
to the DSEs in Fig.~\ref{fig:funDSE}. This is detailed in
Appendix~\ref{app:fun&scale}. It is well-known from the evaluation of
critical physics that the DSE is less amiable to the discussion of
scaling than the FRG. In general scaling in the DSE only comes from a
combination of diagrams which hosts cancellations effectively leading
to the substitution of the bare vertex present in each diagram with
dressed ones. In general such a cancellation also includes the
classical term. One example for such cancellations is the 
$\phi^4$-theory. 

In the present case, however, the functional ghost DSE (second line of
Fig.~\ref{fig:funDSE}) prohibits cancellations between diagrams as it
only consists of one diagram. It is for this reason that the DSE 
system of Yang-Mills theory cannot be subject to cancellations except
for kinematical and global symmetry reasons. Moreover, in both sets of
equations Fig.~\ref{fig:funFRG}, Fig.~\ref{fig:funDSE} we have the
classical term in the DSE or the initial condition for the FRG $\delta
S_{(\Lambda)}/\delta \phi$. For infrared enhanced vertices and inverse
propagators this term is subleading. On the other hand, scaling of
infrared suppressed vertices and inverse propagators requires
cancellations on the right hand side of the FRGs and DSEs between the
diagrams and the classical term. As will become clear later the only
place where such a cancellation necessarily has to occur for a scaling
solution are the DSE and FRG of the ghost propagator. The presence or
absence of cancellations in these equations therefore decides about
the existence of the scaling solution, and is related to global
properties of the gauge fixing, namely the Kugo-Ojima confinement
criterion and the Gribov-Zwanziger horizon condition, see also 
\cite{Fischer:2008uz}.

The functional DSEs and FRGs are derived from the effective action of the 
theory, expanded in its Green functions. With the abbreviation 
\begin{equation}
  \int_{p_1 \cdots p_l} \equiv \int \prod_{i=1}^l 
  \left(\frac{d^d p_i}{(2\pi)^d}\right)
  (2\pi)^d \delta^d\left(\sum_{j=1}^l p_j\right). 
\end{equation}
the effective action is given as 
\begin{eqnarray}\label{eq:G0}
  \Gamma[\phi]&=& \sum_{m,n} \0{1}{m! (n!)^2 } \int_{p_1 \cdots p_{2n+m}}
   \Gamma^{(2n,m)}(p_1 \cdots p_{2n+m})\,\nonumber\\ 
                  &&\hspace*{10mm}
		    \times\prod_{i=1}^n \bar C(p_i) 
                    \prod_{i=n+1}^{2n} C(p_i) 
		    \prod_{i=2n+1}^{2n+m} A(p_i), \nonumber\\
              &\equiv& \sum_{m,n} \0{1}{m! (n!)^2 } 
  \Gamma^{(2n,m)}\, \bar C^n\, C^n\,A^m\,,
\end{eqnarray} 
with the gluon field $A$ and the (anti-)ghost fields $\bar C, C$,
$\phi=(A,\bar C,C)$ 
and suppressed Lorentz- and colour-indices. In the third line in 
Eq.(\ref{eq:G0}) we have introduced an abbreviated 
notation which will be used throughout this work. 
The one-particle-irreducible Greens
functions $\Gamma^{(2n,m)}$ with $2n$ external (anti-)ghost legs and
$m$ external gluon legs are the expansion coefficients of the
effective action in the field expansion. It is convenient to
reparameterise these expansion coefficients with the help of the
coefficients
\begin{eqnarray}
\Gamma^{(2,0)} \equiv Z^{(2,0)}\, S_{\rm cl}^{(2,0)},\qquad  
\Gamma^{(0,2)} \equiv Z^{(0,2)}\, S_{\rm cl}^{(0,2)}
\end{eqnarray} 
of the kinetic terms. We then obtain the
rescaled coefficients $\bar \Gamma^{(2n,m)}(\vec p)$ given by
\begin{eqnarray} \label{eq:sym}
\Gamma^{(2n,m)}(\vec p)
  &=&\,\bar\Gamma^{(2n,m)}(\vec p) \\[1ex] &&
  \times \prod_{i=1}^{2n} \sqrt{Z^{(2,0)}(p_i)} \prod_{i=1}^m
  \sqrt{Z^{(0,2)}(p_{2n+i})}\,. 
 \nonumber\end{eqnarray}
with $\vec p=(p_1,...,p_{2n+m})$. 
This parameterisation implies that the coefficients of the two point Green functions 
\begin{eqnarray} \label{barprop}
\bar\Gamma^{(2,0)}(p)=S_{\rm cl}^{(2,0)}(p)\,,\qquad 
\bar\Gamma^{(0,2)}(p)=S_{\rm cl}^{(0,2)}(p)\,,
\end{eqnarray} 
carry only the canonical momentum dependence of the kinetic terms and
$Z^{(0,2)}$, $Z^{(2,0)}$ account for all quantum corrections. 

The reparametrisation \eq{eq:sym} also entails that the $Z$-factors on
the right hand side of \eq{eq:sym} carry the whole renormalisation
group scaling of the vertex functions $\Gamma^{(2n,m)}$ in terms
of the renormalisation scale $\mu$. Together with
the standard renormalisation group (RG) equation of the theory,
\begin{equation}\label{eq:RG} 
\mu\0{d }{d\mu}\Gamma=0\,, 
\end{equation}
we then learn from  \eq{eq:sym} and \eq{eq:RG} that the expansion
coefficients $\bar \Gamma^{(2n,m)}$ do not depend on $\mu$, i.e.
\begin{equation}\label{eq:barRG} 
\mu\0{d }{d\mu}\bar\Gamma^{(2n,m)}=0\,,\qquad \forall n,m\in \Z\,. 
\end{equation}
Note in this context that \eq{eq:RG} is also valid in the presence of
the RG-adapted regulator terms in the functional renormalisation group
equations (FRGs) \cite{Pawlowski:2005xe,Pawlowski:2001df}. This allows
us to derive constraints for the vertex functions $\bar\Gamma^{(2n,m)}$
also in the presence of the regulator, see
\cite{Pawlowski:2003hq,Fischer:2006vf} for details. In the present
work we also show, see Appendix~\ref{app:fun&scale}, that the
FRG-analysis of the IR-asymptotics can be further simplified, allowing
for a more direct approach.

\Eq{eq:barRG} already suggests that the RG-invariant coefficients
$\bar \Gamma^{(2n,m)}$ do not carry any (global) anomalous scaling in
terms of momenta. We shall show in the following that this is indeed
the case.

\section{Uniqueness of the global scaling}\label{sec:proof}
For this proof we are only interested in the global scaling behaviour
for the coefficient functions $\bar\Gamma^{(2n,m)}$. 
Modulo logarithms this entails the global scaling 
\begin{equation}
\lim_{\lambda\to 0}\bar\Gamma^{(2n,m)}(\lambda \vec p)= 
  \lambda^{2(d_{2n+m}+\bk^{\ }_{2n,m})} \bar\Gamma_{\rm as}^{(2n,m)}(\vec 
p)\,, \label{eq:scalings} 
\end{equation} 
where $\bar\Gamma_{\rm as}$ stands for the
infrared leading term, and  $\vec p=(p_1,...,p_{2n+m})$. The coefficient 
\begin{equation}\label{eq:d2nm}
  d^{\ }_{l}=\0d2 -l\0{d-2}{4}\,, 
\end{equation} 
is the canonical scaling dimension of the vertex $\Gamma^{(2n,m)}$ with 
$l=2n+m$. 
Note that it is only sensitive to the total number of external legs.
It can be directly derived from Eq.~(\ref{eq:G0}): plugged into
Eq.~(\ref{eq:G0}) it matches the canonical scaling of the
$\delta$-function of total momentum conservation, $-d/2$, and that of
ghost, anti-ghost and gluon fields in position space, $(d-2)/4$.  Thus
it matches the momentum scaling of the momentum integral
$\int_{p_1\cdots p_{2n+m}}$ and that of the fields in momentum space.
As a result the scaling (\ref{eq:d2nm}) includes the canonical
momentum scaling of the one-particle irreducible Green functions as
well as the scaling of the couplings. Hence, only in the critical
dimension of Yang-Mills theory, $d=4$, the canonical scaling dimension
(\ref{eq:d2nm}) agrees with the classical momentum scaling. 

It turns out that the present parameterisation (\ref{eq:scalings}) in
terms of $\bk_{2n,m}$ enables us to significantly simplify the proof
given in Ref.~\cite{Fischer:2006vf} of the uniqueness of the
$\kappa_{2n,m}$. At its core the reason is the natural book-keeping of
the necessary RG-scaling by the $Z$-factors that incorporate one
factor of $1/2\kappa_{2,0}$ or $1/2 \kappa_{0,2}$ for each external
leg of the vertex $\Gamma^{(2n,m)}$.

As for the most basic $\bk^{\ }_{2n,m}$ we obtain by definition 
(cp. Eq.~(\ref{barprop}))
\begin{equation}
\bk_{2,0}=\bk_{0,2}=0. \label{barkappaprop}
\end{equation} 
Then the scaling relations for the kinetic terms, the ghost and
gluon dressing functions, read
\begin{eqnarray}\nonumber 
 \lim_{\lambda\to 0}Z^{(2,0)}(\lambda p)&=&
  \lambda^{\kappa^{\ }_{2,0}}Z^{(2,0)}(p) \,, \\[1ex] 
\lim_{\lambda\to 0}Z^{(0,2)}(\lambda p)&=&
  \lambda^{\kappa^{\ }_{0,2}}Z^{(0,2)}(p)\,. 
\label{eq:propscalings} \end{eqnarray}

The total global scaling $\lambda^{2t_{2n,m}}$ of the full vertices
$\Gamma^{(2n,m)}$ also involves the anomalous dimensions of the
propagators and reads
\begin{equation} \label{eq:total} t_{2n,m}=d_{2n+m}+\012
  \left(2n\kappa_{2,0}+m \kappa_{0,2}\right)+\bk_{2n,m}\,. 
\end{equation} 
Previous analyses in
\cite{Alkofer:2004it,Fischer:2006vf,Huber:2007kc,Alkofer:2008jy} were
initiated similarly. However, instead of evaluating the deviation
$\bk_{2n,m}$ to the standard anomalous scaling, the deviation
$\kappa_{2n,m}$ from the canonical scaling in the critical dimension four
of Yang-Mills theory was evaluated, 
\begin{equation} 
t_{2n,m}=d_{2n,m}|_{d=4}+\kappa_{2n,m}\,,
\end{equation}
where
\begin{equation}
  \kappa_{2n,m} = \Delta d_{2n+m} + \frac{1}{2}\left(2n\kappa_{2,0} + 
    m\kappa_{0,2}\right) + \bk_{2n,m} \,, \label{eq:kappabar}
\end{equation}
with the deviation $\Delta d_{2n+m}$ of the 
canonical scaling from that in the critical dimension four, 
\begin{equation}\label{eq:Deltad2nm}
  \Delta d^{\ }_{l} = d^{\ }_{l}-d^{\ }_{l}|_{d=4}
  = (2-l) \frac{d-4}{4}\,, 
\end{equation}
and $l=2n+m$. This is adapted such that the $\kappa_{2n,m}$ for
primitively divergent vertices describe the full scaling of the
corresponding dressing functions in $d$ dimensions.

\begin{figure*}
\begin{center}
\includegraphics[width=0.8\textwidth]{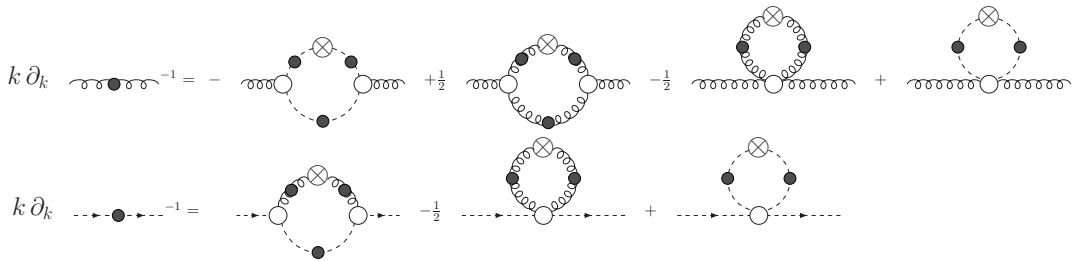}
 \caption{Functional renormalisation group equations for the gluon and
   ghost propagator.
   Filled circles denote dressed propagators and
   empty circles denote dressed vertex functions. Crosses indicate
   insertions of the infrared cutoff function. Only
   one possible insertion of the infrared cutoff function per diagram
   is shown.}
 \label{fig:ERGE-diag}
\end{center}
\end{figure*}

We emphasise that in principle additional logarithmic scalings should
be included into \eq{eq:scalings},\eq{eq:propscalings}. However, even
if present, additional logarithmic scalings do not change the
relations between the $\bar\kappa_{n,m}$ and are therefore irrelevant
for the purpose of the present investigation. We also add that
self-consistent logarithmic scaling laws have to satisfy additional
consistency conditions.

In four dimensions an explicit solution of the whole tower of DSEs and 
FRGs in terms of $\kappa_{2n,m}$ was first derived in \cite{Alkofer:2004it} 
and then generalised to $d$ dimensions in \cite{Huber:2007kc}. It reads 
\begin{eqnarray}
\kappa &\equiv& \kappa_{2,0} = -\0{4-d}{4}-\012\kappa_{0,2}\nonumber\\[1ex]
\kappa_{2n,m} &= &(n-m)\kappa+(1-n)\left(\frac{d}{2}-2\right) \,.  
\label{eq:previous}
\end{eqnarray}
In \cite{Fischer:2006vf} we already gave a proof for the uniqueness of
Eq.~(\ref{eq:previous}). In the following we reformulate this proof in
a, to our mind, more elegant and instructive manner that also allows
us to reduce the number of presuppositions. We first derive
constraints for $\bk_{2n,m}$ from the functional renormalisation group
and Dyson-Schwinger equations and then show that $\bk_{2n,m}=0$ for
all $n,m$. The resulting expression for the $\kappa_{2n,m}$ from
Eq.~(\ref{eq:kappabar}) then agrees with Eq.~(\ref{eq:previous}).

\subsection{Constraints from the functional RG}\label{sec:relFRG}

The FRGs for the ghost and gluon propagators are given
diagrammatically in Fig.~\ref{fig:ERGE-diag}. With a mode cut-off,
which only removes a single momentum mode, the regulator insertion is
proportional to a $\delta$-function and simply restricts the loop
integral to a given momentum $p^2$ which we take to be vanishing.
Then the loops on the rhs of the flow simply count the powers of
global momentum scaling of the quantum fluctuations, no initial
condition, similar to the classical term in the DSE, appears, see also
\cite{Fischer:2006vf}. The potential cancellations necessary for the
initial condition are discussed at the end of our proof.

We are now counting anomalous dimensions on both sides of the
equations in terms of powers of one external momentum scale $p^2$ in
the infrared region $p^2 \ll \Lambda_{QCD}^2$.  For the global 
scaling (\ref{eq:scalings}) considered here all anomalous dimensions
in terms of internal momenta of the loops translate directly into
anomalous dimensions of the external momentum scale. This is also true
for the vertex equations considered below. In this respect the
regulator insertion, denoted by the crosses, carry the anomalous
dimensions of inverse propagators
\cite{Pawlowski:2003hq,Fischer:2006vf}.

The constraint equations for $\bar\kappa_{2n,m}$ can be derived in
several ways. A somewhat pedestrian approach is to count anomalous
dimensions $\kappa_{2n,m}$ of the dressing functions on both sides 
of the equations and then
converting to $\bk_{2n,m}$ with the help of (\ref{eq:kappabar}).  More
efficiently, we note that the $\kappa_{2,0},\kappa_{0,2}$ carry the
renormalisation group scaling of the corresponding Green functions and
match on both sides of the FRG equations. In particular this is true
for the propagator FRGs in Fig.~\ref{fig:ERGE-diag}.  Consequently all
$\kappa_{2,0},\kappa_{0,2}$ drop out of the FRG-relations for a
general vertex $\Gamma_{2n,m}$. Note also that the sum of the
canonical dimensions $d_{2i,j}$, \eq{eq:d2nm}, in a given diagram for
$\Gamma_{2n,m}$ simply gives the total canonical dimension $d_{2n+m}$,
and hence the $d_{2i,j}$ also drop out of the FRG-relations.  Then, we
are left with relations for solely the $\bar\kappa_{2i,j}$. 
For the propagators the constraints read 
\begin{eqnarray}
  0 = \bk_{0,2} &=& \mbox{min}\left(2 \bk_{2,1}\,,\, 2 
    \bk_{0,3}\,, 
    \,\bk_{0,4}\,, \,
    \bk_{2,2}\right)\,,\label{eq:glue-frg}\\[1ex]
  0 = \bk_{2,0} &=& \mbox{min}\left(2 \bk_{2,1}\,,\, 
    \bk_{2,2}\,,\, \bk_{4,0}
  \right)\,,
\label{eq:ghost-frg}
\end{eqnarray}
from the gluon and ghost FRGs. For the lhs of these equations
we used that $\bk_{2,0}=\bk_{0,2}=0$ by definition,
cp. Eq.(\ref{barkappaprop}).
The minimum prescription on the right hand side of \eq{eq:glue-frg},
\eq{eq:ghost-frg} takes into account that only one of the diagrams may
be leading in the infrared. The constraint (\ref{eq:ghost-frg}) from the
ghost-FRG entails
\begin{eqnarray}
  \bk_{2,1} \ge 0\,,\qquad \qquad  \bk_{2,2} \ge 0\,, \qquad \qquad 
  \bk_{4,0} \ge 0\,, \label{eq:con1}
\end{eqnarray}
and at least one of these has to be zero for \eq{eq:ghost-frg} to be satisfied,
\begin{eqnarray}
  \bk_{2,1} =0\,,\qquad {\rm or} \qquad  \bk_{2,2} = 0\,, \qquad {\rm or} 
  \qquad 
  \bk_{4,0} = 0\,. \label{eq:con1b}
\end{eqnarray}
The same analysis for \eq{eq:glue-frg} entails that
$\bk_{2,1},\bk_{0,3},\bk_{0,4},\bk_{2,2}\geq 0$ with at least one of them being
zero. For the proof below, however, Eqs.~(\ref{eq:con1}) and (\ref{eq:con1b}) 
will be sufficient.

\begin{figure*}
\begin{center}
\includegraphics[width=0.8\textwidth]{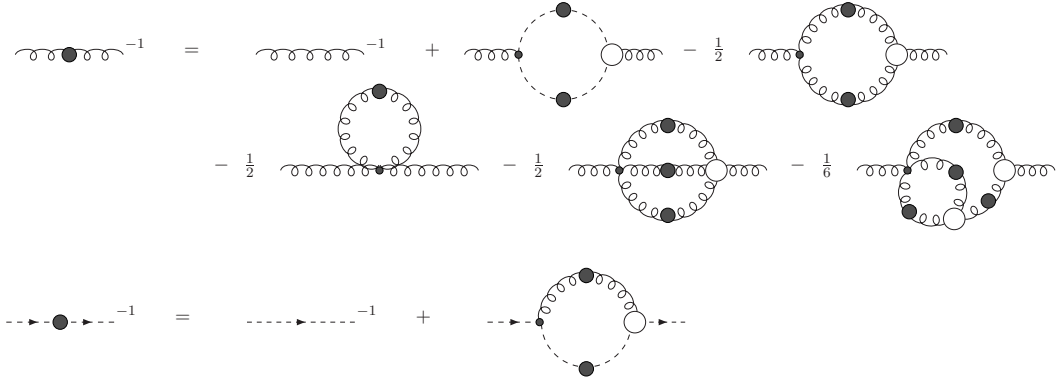}
 \caption{Dyson-Schwinger equations for the gluon and
   ghost propagator. Filled circles denote dressed propagators and
   empty circles denote dressed vertex functions.}
 \label{fig:DSE-diag}
\end{center}
\end{figure*}

We conclude the FRG-analysis with a discussion of the FRG relations for 
general Green functions. Schematically these relations read 
\begin{equation}\label{eq:genFRG}
\bk_{2n,m}=\min\left( \bk_{2n+2,m}\,,\,  \bk_{2n,m+2}\,,\,
\dots  \right)\,, 
\end{equation}
where the first two terms are the tadpole contribution with ghost
tadpole ($\bk_{2n+2,m}$), and a gluon tadpole ($\bk_{2n,m+2}$)
respectively. The dots stand for other diagrams with at least two
vertices. It follows that $\bk_{2n,m}$ appears as the tadpole
contribution in the relation for $\bk_{2n-2,m}$ and $\bk_{2n,m-2}$,
and more generally
\begin{equation}\label{eq:geninequality}
\bk_{2(n-r),m-2s}\leq \bk_{2n,m}\,,
\end{equation}
for all $r<n$ and $2s < m$.  This allows us to relate general
$\bk_{2n,m}$ to either $\bk_{2,1}$ for odd $m$ and $2s=m-1$, or
$\bk_{2,2}$ for even $m$ and $2s=m-2$. Thus we have 
\begin{equation}\label{eq:fininequality} 
\bk_{2n,m}\geq \left\{
\begin{array}{ll} \bk_{2,2} & {\rm for}\ m\ {\rm even} \\[1ex] 
\bk_{2,1} & {\rm for}\ m\ {\rm odd} \end{array} \right.\,,  
\end{equation} 
and we conclude with \eq{eq:con1} that  
\begin{equation} \label{eq:geq0}
\bar \kappa_{2n,m}\geq 0\,,\qquad \qquad \forall n,m\in \N\,,
\end{equation}
in general space-time dimension $d$. This constraint together with
\Eq{eq:con1b} will be important in what follows, as it summarises in a
closed form the infinite number of constraints from higher diagrams.

\subsection{Constraints from Dyson-Schwinger equations}

\begin{figure*}
\begin{center}
\includegraphics[width=0.8\textwidth]{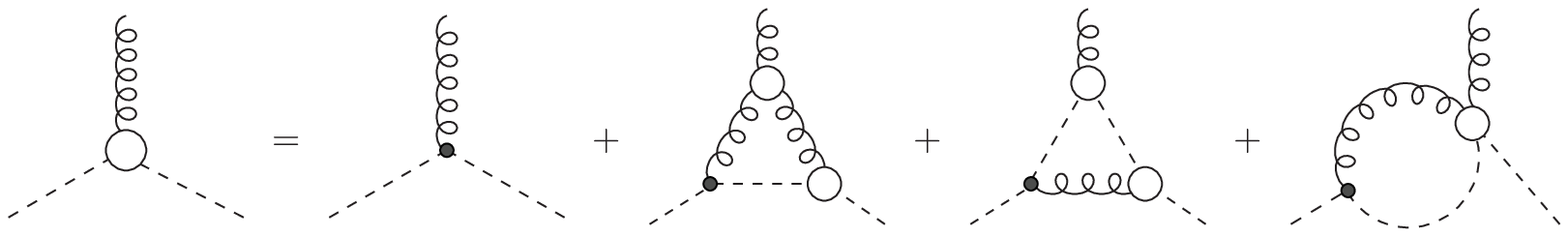}
\includegraphics[width=0.8\textwidth]{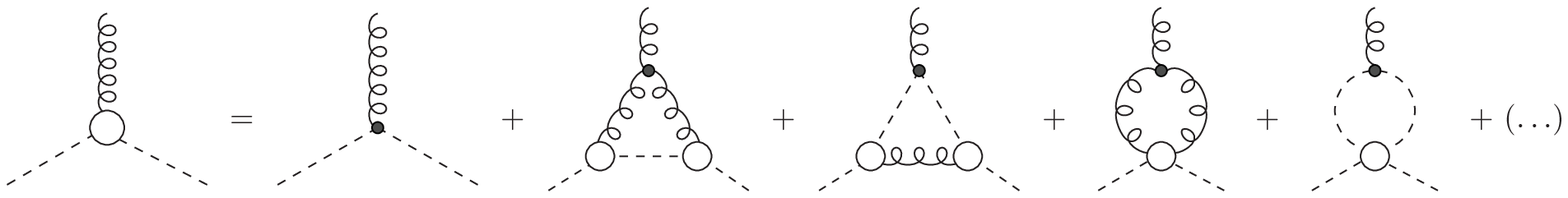}
\caption{Dyson-Schwinger equations for the ghost gluon vertex. Empty
  circles denote dressed vertex functions. All internal propagators
  are dressed; the corresponding filled circles have been omitted for
  clarity of the figures. One-loop diagrams with the same scaling
  behaviour are only shown once.  The ellipses denotes the other one-
  and two-loop diagrams which are not needed for our analysis.}
 \label{fig:DSE-vert}
\end{center}
\vspace*{-2mm}
\end{figure*}

The Dyson-Schwinger equations for the ghost and gluon propagators are
given diagrammatically in Fig.~\ref{fig:DSE-diag}, whereas the
corresponding equations for the ghost-gluon vertex are displayed in
Fig.~\ref{fig:DSE-vert}.  For the ghost gluon vertex we have two DSEs
which are derived from either the functional gluon DSE or the
functional ghost DSE, see \cite{Fischer:2006vf}. As already mentioned,
the potential cancellations necessary for the classical terms are
discussed at the end of our proof.

We have seen in the analysis of the FRGs that the
$\bar\kappa_{2n,m}$-constraints boil down to simply counting the
vertices involved in a given diagram and summing up the corresponding
$\bk_{2n,m}$.  The same would apply to the DSEs if we only had dressed
vertices in the DSE diagrams. However, there is always one bare vertex
which then counts as $\bk_{2n,m}-\kappa_{2n,m} \equiv
-\Delta\kappa_{2n,m}$.  These differences are given by
\begin{equation}\label{eq:dk2nm}
  \Delta\kappa_{2n,m} = \Delta d_{2n+m}
+\012\left(2n\kappa_{2,0}+m\kappa_{0,2}\right) \,. 
\end{equation}
For example, we are thus led to $\bk_{2,1}-\Delta\kappa_{2,1}$ for 
the right hand side of the ghost propagator DSE, and zero on the 
left hand side similar to the ghost FRG. This simple
counting applies to all the diagrams.
For its chief importance in the proof we introduce the abbreviation
\begin{equation}\label{eq:dk}
  \Delta\kappa \equiv \Delta\kappa_{2,1}.
\end{equation}
The constraints derived from the propagator DSEs displayed in 
Fig.~\ref{fig:DSE-diag} are then given by
\begin{eqnarray}\nonumber 
 0 &=& \mbox{min}\Bigl(\bk_{2,1}-\dk\,,\, \bk_{0,3}-\dk_{0,3}\,,\,
-\dk_{0,4}\,,\, \hspace{1cm}\\[1ex]
    & & \hspace{2cm}\bk_{0,4}-\dk_{0,4}\,,\,
    2\bk_{0,3}-\dk_{0,4}\Bigr)\,, \label{eq:congluon}\\[1ex]
  0 &=& \mbox{min}\left(\bk_{2,1}-\dk\right)\,. \label{eq:ghost}
\end{eqnarray}
Certainly, these relations can be derived as well in the pedestrian way of
counting $\kappa_{2n,m}$ on both sides of the equations and converting
them to $\bk_{2n,m}$. Note that in contradistinction to the FRG equations 
the DSEs do depend on $\kappa_{2,0}$ and $\kappa_{0,2}$ via the
$\Delta\kappa_{2n,m}$.

In the two different DSEs for the ghost gluon vertex in
Fig.~\ref{fig:DSE-vert} we again apply the now familiar 
counting and obtain
\begin{equation}
  \bk_{2,1} = \mbox{min}\left(\bk_{2,1}+
    \bk_{0,3}-\dk\,,\, 2\bk_{2,1}-\dk\,,\,\bk_{2,2}-\dk\right)\,, 
  \label{eq:ghost1}\end{equation}
from the upper relation in Fig.~\ref{fig:DSE-vert} and
\begin{eqnarray}\nonumber  
\hspace{-.5cm}  \bk_{2,1} &=& \mbox{min}\Bigl(2\bk_{2,1}-
    \dk_{0,3}\,,\, 2\bk_{2,1}-\dk\,,\, \\[1ex]
    && \hspace{1cm}\bk_{2,2}-\dk_{0,3}\,,\,
    \bk_{4,0}-\dk\,,\, \mbox{two-loop} \Bigr)\,, \label{eq:ghost2}
\end{eqnarray}
from the lower relation in Fig.~\ref{fig:DSE-vert}. These constraints
will be used in the next subsection. We emphasise that the above relations 
are valid in arbitrary dimensions as the FRG relations derived in the 
last Section~\ref{sec:relFRG}. In contradistinction to the FRG-relations 
the DSE relations depend on the dimension via the $\dk_{2n,m}$.

\subsection{Proof}
We now proceed to show that
\begin{equation}
  \bk_{2n,m}=0
\end{equation}
is the only scaling solution of infrared Yang-Mills theory in Landau
gauge. To this end we note that scaling in the ghost-DSE (\ref{eq:ghost}) 
implies that 
\begin{equation}
  \dk = \bk_{2,1} \ge 0\,, \label{eq:dk-eq}
\end{equation}
where we have
used that the ghost-FRG entails $\bk_{2,1}\geq 0$. Furthermore we
obtain the two constraints
\begin{eqnarray}
  \bk_{4,0} \ge \dk\,,\qquad 
\qquad \bk_{2,2}\geq \dk\,,
 \label{eq:con3}
\end{eqnarray}
from the two DSEs for the ghost gluon vertex, where we use 
$0 \le \bk_{2,1}\leq \bk_{2,2}-\dk$ from (\ref{eq:ghost1}), and
$0 \le \bk_{2,1} \le \bk_{4,0}-\dk$ from (\ref{eq:ghost2}). 
However, the ghost FRG led to the 
constraint \eq{eq:con1b}. Together with (\ref{eq:dk-eq}) and
(\ref{eq:con3}) this immediately leads to
\begin{equation}\label{eq:dk=0}
\dk=0\,,
\end{equation}
and therefore also $\bk_{2,1} = 0$ due to (\ref{eq:dk-eq}). 

It remains to be shown that this implies that all of the $\bk_{2n,m}$
need to be zero. To this end we resort to the FRG-equations. The
FRG-equations for all $\bk_{2n,m}$ contain at least one diagram that
solely depends on $N_{2n,m}$ ghost-gluon vertices.  This implies
\begin{equation}\label{eq:0nm}
\bk_{2n,m}\leq N_{2n,m}\bk_{2,1} =0\,.
\end{equation} 
Together with \eq{eq:geq0} this leads to  
\begin{equation}
  \bk_{2n,m} =  0\,.\label{eq:con6}
\end{equation}
Written in terms of our original anomalous dimensions $\kappa_{2n,m}$,
see (\ref{eq:kappabar}), this implies that
\begin{equation}\label{eq:scaling}
  \kappa_{2n,m}= \Delta d_{2n+m}+
\frac{1}{2}\left(2n\kappa_{2,0} + m\kappa_{0,2}\right)\,,
\end{equation}
with 
\begin{equation}\label{eq:scalingDeltad}
  \Delta d_{2n+m} = (2-2n-m) \frac{d-4}{4}\,,  
\end{equation}
see \eq{eq:kappabar} and \eq{eq:Deltad2nm} respectively.
\Eq{eq:scaling} and \eq{eq:scalingDeltad} represent the 
unique scaling solution of infrared Yang-Mills theory.

However, we would like to emphasise that the values of the anomalous
scalings $\kappa_{2,0},\kappa_{0,2}$ of the propagators cannot be
fixed by scaling arguments. They have to be computed by solving the
corresponding FRG and DSE equations, see e.g.\
\cite{Fischer:2008uz,Pawlowski:2003hq,Lerche:2002ep,Zwanziger:2001kw,%
  Fischer:2002hna}. Instead, there is one final piece of information
which can be extracted from the scaling analysis, namely constraints
on the values for $\kappa_{2,0},\kappa_{0,2}$: From \eq{eq:congluon}
we deduce $\dk_{0,4}\le 0$. Using \eq{eq:dk2nm} we have
$(4-d)/2+2\kappa_{0,2}\leq 0$, and with \eq{eq:dk=0} we then conclude
that \cite{Zwanziger:2001kw,Maas:2004se}
\begin{equation}\label{eq:kappa>0}
\kappa_{2,0} =-\0{4-d}{4}-\012\kappa_{0,2}\ge -\0{1}{2} \0{4-d}{4}\,.  
\end{equation}
\Eq{eq:kappa>0} is in accordance with the Gribov-Zwanziger confinement
scenario \cite{Gribov,gzwanziger2} which predicts ghost-enhancement.

Finally we discuss the fate of the classical terms with
$\kappa_{0,2}^{\rm class}=\kappa_{2,0}^{\rm class}=0$. The propagator
scaling stemming from the quantum fluctuations is summarised
qualitatively as
\begin{equation}\label{eq:<>}
  \kappa_{2,0}\geq -\0{1}{2} \0{4-d}{4}
  \,,\qquad \qquad \kappa_{0,2} \leq -\0{4-d}{4}\,.
\end{equation}
With $\kappa_{0,2}\leq \kappa_{0,2}^{\rm class}=0$, the classical 
contribution does not change the gluon scaling.  In turn, for 
$\kappa_{2,0}>0$, the inverse ghost propagator would be dominated 
by its classical part, which has to be cancelled if scaling applies. 
This is the adjustment of the horizon condition, see 
e.g.\ \cite{Lerche:2002ep,Fischer:2008uz}. Note also that the 
bound \eq{eq:<>} in principle also allows for $\kappa_{2,0}<0$ for $d<4$. 
In practice, however, one finds $\kappa_{2,0}>0$, i.e. 
$\kappa_{2,0}(d=3)= 0.40$, and $\kappa_{2,0}(d=2)= 1/5$ for a 
classical ghost-gluon vertex \cite{Zwanziger:2001kw}. 
For these values to drop below zero, the full ghost-gluon vertex 
would have to deviate drastically from the classical vertex.
This is almost excluded by lattice computations 
\cite{Ilgenfritz:2006he,Cucchieri:2008qm} and DSE self-consistency checks 
\cite{Lerche:2002ep,Schleifenbaum:2004id}. 

Numerical solutions for the ghost and gluon DSE as well as the
corresponding FRGs in agreement with \Eq{eq:scaling} and
\eq{eq:scalingDeltad} have been given in
\cite{Fischer:2002hna,Fischer:2008uz,Schleifenbaum:2004id,%
  Fischer:2004uk,Kellermann:2008iw}. In \cite{Bloch:2003yu} a
truncation has been employed which effectively converts the DSEs into
equations with dressed vertices resembling the structure of FRGs. This
procedure allows for interesting numerical solutions. However, the
vertices used are neither consistent with the (unique) infrared
scaling laws \Eq{eq:scaling} and \eq{eq:scalingDeltad}, nor with the
standard RG-scaling. At its core this is due to the fact that the
system of FRG equation cannot be solved by in the IR as the gluonic
vertices in \cite{Bloch:2003yu} are too singular.

These comments complete our proof.
\begin{figure*}
\begin{center}
  \includegraphics[width=0.8\textwidth]{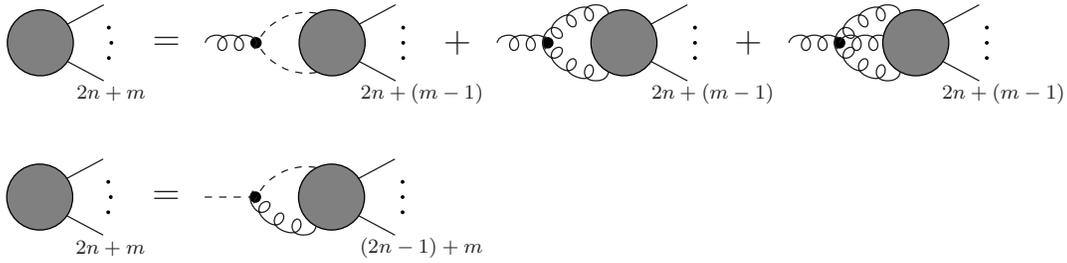}
 \caption{General Dyson-Schwinger equations for $\Gamma^{(2n,m)}$. 
 The straight legs denote either ghost or gluon lines with multiplicity given
 below the diagram, respectively.}
 \label{fig:genDSE-vert}
\end{center}
\end{figure*}

\section{Kinematics} \label{sec:kinematics}

We have derived the unique scaling solution \eq{eq:scaling} with the
assumption of conformal scaling. As already argued in
\cite{Fischer:2006vf} and in the introduction, possible conformal
invariance entails that no further constraint on $\kappa_{2,0}$ and
$\kappa_{0,2}$ apart from \eq{eq:<>} can be derived from pure
scaling arguments. Here we shall make this formal argument more
explicit, and also discuss the question of existence of a specific
scaling solution including various kinematical limits. 

This is also interesting for the following reason: From our analysis
in the previous section we know that the classical purely gluonic
vertices are sub-dominant, and the ghost-gluon vertex has its
classical global scaling. Note however that a ghost gluon vertex with
$\bar\Gamma^{(2,1)}\simeq S^{(2,1)}$ requires the cancellation of the
classical term in the DSE and the initial condition in the FRG
equations in some kinematical limits, i.e.\ vanishing ghost or
anti-ghost momentum. In these limits the diagrammatical scaling would
then be dominated by the classical part/initial condition. 

As already discussed above, the FRG equations cannot provide
constraints on $\kappa_{2,0}$, $\kappa_{0,2}$, as the $\kappa$'s drop
out of the FRG-scaling analysis after employing the parameterisation
\eq{eq:sym}. The FRG-diagrams then reduce to one loop diagrams with
bare propagators and dressed vertices $\bar\Gamma^{(2n,m)}$ with only
canonical global scaling due to (\ref{eq:con6}).  Here and in the
following we use the term 'global scaling' when all external momenta
scale in contrast to 'kinematical scaling' when only some external
momenta are involved\footnote{This corresponds to 'uniform' vs.
  'collinear' singularities in the terminology of
  \cite{Alkofer:2008jy}.}. Note that the $\bar\Gamma^{(2n,m)}$ carry
kinematical singularities as is well-known from perturbation theory.
In Landau gauge these may be meliorated by transversality.  Since we
are working in Landau gauge we are only interested in transversal
contributions to all vertices; in Landau gauge the set of fully
transversal vertices together with the ghost propagator and
transversal gluon propagator provide a closed set of FRG and DSE
equations. In turn, vertices with at least one longitudinal gluon
satisfy DSEs and FRGs that also depend on the transversal vertices and
only on the transversal gluon propagator, see also
\cite{Fischer:2008uz}.

For these reasons we concentrate on the purely transversal part of the
vertices, ${\Gamma_\bot} ^{(2n,m)}$. For the treatment of kinematical
scaling we also extend our notation to
\begin{eqnarray}\nonumber 
  &&\hspace{-.7cm} 
  {\Gamma_\bot}^{(2n,m)}(\lambda \vec p_{1,r},\vec p_{r+1,2n}\,,\,
  \lambda \vec p_{2n+1,2n+s}\,,\,\vec p_{2n+s+1,2n+m})\\[1ex]
  && \stackrel{\lambda\to 0}{
    \longrightarrow} 
  \lambda^{2(d_{r+s}+\kappa_{2n,m}^{r,s})} 
  \,{\Gamma_\bot}_{r,s}^{(2n,m)}(\vec p)\,,  
\label{eq:kinematical}\end{eqnarray}
where ${\Gamma_\bot}_{r,s}^{(2n,m)}$ stands for the infrared leading
term, $\vec p_{i,j}=(p_i,...,p_j)$, and $\vec p$ is not exceptional by
itself. Here, $s$ counts scaling gluon momenta, and $r=r_1+r_2$ counts
$r_1$ scaling ghost momenta and $r_2$ scaling anti-ghost
momenta. Analogously to \eq{eq:total} the total scaling $\lambda^{2
  t^{r,s}_{2n,m}}$ of the full vertex $\Gamma^{2n,m}$ reads
\begin{equation}\label{eq:trs2nm}
t^{r,s}_{2n,m}=d_{r+s}+\kappa_{2n,m}^{r,s}\,. 
\end{equation}
The global scaling (\ref{eq:scaling}) is reproduced by
\begin{equation}
\kappa_{2n,m}^{r,s}=\kappa_{2n,m}\,,\quad \forall r+s=2n+m-1\,,
\end{equation} 
due to momentum conservation.

In the following we shall show that a dressed ghost-gluon vertex
without kinematic singularities is a possible solution of the DSE and
FRG systems. In this case the scaling relations for the ghost-gluon
vertex ${\Gamma_\bot}^{(2,1)}$ are
\begin{equation}\label{eq:ghostgluon}
\kappa_{2,1}^{1,0}=\kappa_{2,1}^{0,1}=\kappa_{2,1}=0\,.
\end{equation}
A ghost-gluon vertex with \eq{eq:ghostgluon} reads 
\begin{equation} \label{eq:barcAc}
{\Gamma_\bot}_{\mu,abc}^{(2,1)}(p,q)
=\Pi^\bot_{\mu\nu}(p) q_\nu f(p+q,q)f^{abc}\,. 
\end{equation} 
where 
\begin{equation}\label{eq:transverse} 
\Pi^\bot_{\mu\nu}(p)=\delta_{\mu\nu}-\0{p_\mu p_\nu}{p^2}\,, 
\end{equation}
and $f$ is a non-singular function of both momenta. Here $p$ is the
gluon momentum, $q+p$ is the ghost momentum and $q$ is the anti-ghost
momentum respectively. Ghost--anti-ghost symmetry is implicit with
\begin{equation}\label{eq:barccsym}
\Pi^\bot_{\mu\nu}(p) q_\nu=\Pi^\bot_{\mu\nu}(p) (p+q)_\mu\,,\quad 
f(p+q,q)=f(q,p+q)\,.
\end{equation}
Now we assume that the leading infrared parts of {\it all} vertices
are already provided by diagrams only depending on ghost-gluon
vertices. This entails that, with the exception of the ghost-gluon
vertex, {\it all} ghost or anti-ghost legs of arbitrary vertices are
proportional to linear powers of the corresponding external momentum. 
The reason is that each of these legs is attached to an internal
transversal gluon. With \eq{eq:barccsym} we then can always rewrite 
this diagram as proportional to linear powers of the external momenta. 

With this fact in mind we can deduce the scaling from the general form
of the vertex-DSEs. Diagrammatically they are depicted in
Fig.~\ref{fig:genDSE-vert}. We begin our analysis with vertices where
one external gluon momentum is vanishing. To this end we take the
external momentum $p=p_{\rm bare}$ of the bare vertex to zero by
keeping the other momenta at non-exceptional values. Extracting the
linear momenta assigned to the internal ghost lines and projecting
onto transverse components we get
\begin{eqnarray}\nonumber 
  &&  \hspace{-0.8cm}
  \Pi^\bot_{\mu\nu}(p)\, \int d q q^{d-1} \int d\Omega_q\, 
  \0{1}{q^{2(1+\kappa)} } 
  \0{1}{(q+p)^{2(1+\kappa)}} 
  \\[1ex]
  && \hspace{-.5cm}  \times \,q_\nu \,(q+p)_\rho\, 
  q_\sigma \,{\cal I}_{\rho\sigma\mu_1 \cdots \mu_{m-1}}(p,q,p_1\,\cdots 
  p_{2n+m-1})\,,  \label{bb}
\end{eqnarray}
where $ \int d\Omega_q$ stands for the angular integration stemming
from the $q$-integration, and $\mu=\mu_m$. 

\begin{figure}[t]
\begin{center}
  \includegraphics[width=0.5\columnwidth]{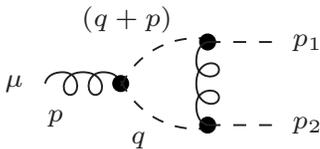}
 \caption{Infrared leading diagram in the DSE for the ghost-gluon vertex.}
 \label{ghost-glue}
\end{center}
\end{figure}

As an example we discuss the simplest possible diagram for the case of
no further external gluons, $m=1$, shown in Fig.~\ref{ghost-glue} with a 
similar momentum routing as in Eq.~(\ref{bb}).
In this diagram each of the three ghost-gluon vertices generates one
four-momentum. Due to the transversality of the involved two gluon
lines these can be converted into the factor $q_\nu \,(q+p)_\rho\,
q_\sigma$ also appearing in Eq.~(\ref{bb}).  For a non-vanishing
integral we need a further power in $q$, otherwise the integral is
proportional to $p_\nu$ and vanishes due to projection with
$\Pi^\bot_{\mu\nu}(p)$. Indeed an extra factor $p_i\cdot q$ for some
$i=1,...,2$ is generated by the internal gluon. In the general
expression (\ref{bb}) this extra factor is provided by the kernel
${\cal I}$.  Counting powers of the scaling momenta $p$ and $q$ we
find
\begin{equation} \label{eq:01ghost}
\kappa_{2n,1}^{0,1}=  \frac{d}{2}-2\kappa\,, 
\end{equation}
for $m=1$. From (\ref{bb}) we have $d/2$ from the integration, $-2$
from the two denominators of the propagators, $-2\kappa$ from the two
ghost dressing functions in the propagators and $+2$ from the four
powers of $q$. The other momenta, $p_1$ and $p_2$ do not scale. The
meaning of Eq.(\ref{eq:01ghost}) is the following: we only obtain an
infrared dominated, divergent and therefore scaling integral for
$\kappa>d/4$, otherwise we cannot say something definite from the
scaling analysis. Whether the integral is then vanishing or finite
depends on the details of the kernel ${\cal I}$ and the angular
integration; consequently the integral may or may not display
kinematical scaling for vanishing gluon momentum in this case.

On the other hand, in the general case where further external gluons
are present, i.e. $m>1$, we have contractions of $q_\nu (q+p)_\rho
q_\sigma$ with ${\cal I}_{\rho\sigma\mu_2 \cdots
  \mu_{m}}(p,q,l_2\,\cdots l_{2n+m})$.  These contractions generate
terms proportional to $p_{\mu_i}$ with $i = 1,...,(m-1)$ that do not
vanish even when all external gluon legs are transversally projected.
Thus we can have terms in the integrand proportional to $q_\nu p_\rho
q_\sigma$. In order to decide whether the integral produces
kinematical singularities or not we count powers in scaling loop
momenta, i.e.  two factors of $q$, two ghost dressing functions, two
denominators and the integration. We arrive at the condition
\begin{equation} \label{eq:01glue_cond}
\frac{d-2}{2}-2\kappa<0 \qquad \Rightarrow \qquad \kappa>\frac{d-2}{4} \,.
\end{equation}
Thus kinematical singularities only occur for $\kappa >
  (d-2)/4$.  However, for the scaling of the vertex in terms of the
external gluon momentum we have to count all powers of scaling momenta
$q$ and $p$ and arrive at\footnote{Since we also count powers of
  $p_{\mu_i}$ this relation possibly involves the tensor structure of
  the vertex $\Gamma_{2n,m}$. The only case where this is important
  and interesting is the special case of the three-gluon vertex
  $\Gamma_{0,3}$. It is easy to see that the tensor structure of the
  bare three gluon vertex cannot have kinematical singularities and
  therefore cannot take part in kinematical scaling. Since
  Eq.~(\ref{eq:01glue}), however, involves an explicit scaling
  momentum with external Lorentz-index it has to represent the
  kinematical scaling of a different tensor structure of the
  three-gluon vertex.}
\begin{equation} \label{eq:01glue} \kappa_{2n,m}^{0,1}=
  \frac{d-1}{2}-2\kappa\,,
\end{equation}
for $m>1$, in agreement with \cite{Alkofer:2008jy} for $d=4$.

From Eqs.~(\ref{eq:01ghost}) and (\ref{eq:01glue}) we observe that the
kinematical gluonic singularities are smaller or equal to the lowest
possible global singularity $-\kappa$ as long as $\kappa\leq
  (d-1)/2$, i.e.
\begin{equation} \label{eq:lessthanglobal}
\kappa_{2n,m}^{0,1}\geq -\kappa \quad \rightarrow \quad \kappa\leq \frac{d-1}{2}
\end{equation}
Consequently the gluonic vertex dressing in the scattering kernels in
Fig.~\ref{fig:genDSE-vert} does not lead to divergences for $p\to 0$
in neither the gluonic nor the ghost DSEs. In particular this entails 
that 
\begin{equation} \label{eq:10ghost}
\kappa_{2n,m}^{1,0}\geq 0\,. 
\end{equation}

We conclude our analysis with a brief discussion of higher kinematical
singularities, which is worked out in more detail in Appendix
\ref{app:vertices}. There we derive the scaling relation
\begin{equation}\label{eq:scalingsum}
t^{r,s}_{2n,m}=
t_{r,s}+\left\{\begin{array}{lc} \displaystyle 
-\0{d-2}{4}
+\012 \kappa \qquad 
& \displaystyle  r_2= r_1+1\,,\\[2ex]
\displaystyle \min\left(-\kappa, \frac{2-d}{2}+2\kappa \right) \qquad 
& \displaystyle {\rm else} \end{array}\right. 
\end{equation}

To summarise: for $\kappa\leq (d-1)/2$ we have obtained kinematical
divergencies that are small enough such that they cannot invalidate
the global scaling relation \eq{eq:scaling}. Note however that the
bound for the existence of a Fourier transform of the ghost propagator
is $\kappa< (d-2)/2$, see also \cite{Lerche:2002ep}.  If $\kappa$
exceeds $(d-2)/2$, the ghost propagator in position space cannot be
understood anymore as tempered distributions which is a necessary
condition for correlation functions in a local quantum field
theory. Hence the above bound $\kappa\leq (d-1)/2$ is not relevant for
this physically interesting case. We conclude that the relation
\eq{eq:scalingsum} for the kinematical scaling together with the
global scaling Eq.~(\ref{eq:scaling}) is a possible solution of the
tower of DSEs and FRGs.

We do not want to further this discussion, in particular as the actual
numerical -and physically sensible- solution satisfy all of the above
bounds, that is $\kappa_{\rm num}<(d-2)/2$. We merely would like to
mention that even in the case $\kappa>(d-1)/2$ one cannot conclude
that the system is ill-defined. It only means that such an anomalous
scaling cannot be captured by the DSE tower of equations, where the single
diagrams do not entail the full RG-scaling, and hence also do not have
manifest scale invariance in the scaling region.  A consistent
solution to such a system necessarily requires non-trivial
cancellations between different diagrams. In the ghost propagator DSE,
where such cancellations cannot occur, they are not necessary; the
ghost-DSE is well-defined even for $\kappa>(d-1)/2$ in the case of
full kinematical scaling of the ghost-gluon vertex.

From our kinematical analysis we find that in the case of 
$\kappa>(d-2)/2$ all vertices have kinematic singularities. 
As stated above the DSEs cannot be used anymore. The FRGs, however, 
can straightaway be solved by $\bk^{r,s}_{2n,m}=0$.  This
is also the reason why the above solution, Eq.~(\ref{eq:scalingsum}),
for kinematical scaling is not unique in contradistinction to the
global scaling relation Eq.~(\ref{eq:scaling}). These comments
complete our proof of existence.

\section{Summary}

In this work we demonstrated that there is only one, unique global
scaling solution of infrared Yang-Mills theory in Landau gauge. To
this end we introduced a parameterisation of the one-particle
irreducible Green functions of the theory that splits off the
renormalisation group (RG)-scaling and reduces the discussion to that
of the scaling properties of RG-invariant quantities. To our mind this
greatly simplifies the proof as compared to our previous work
Ref.~\cite{Fischer:2006vf}. We wish to emphasise again, that the
values of the anomalous scalings $\kappa_{2,0},\kappa_{0,2}$ of the
propagators cannot be fixed by scaling arguments. These have to be
calculated explicitly from the corresponding FRG and DSE equations.

In addition we also demonstrated the existence of a specific scaling
solution that includes 'kinematical scaling' in various kinematical
limits. In contradistinction to the global scaling relation,
kinematical scaling is not unique.

In general, the method presented here is also applicable to other theories.
As an example we shortly discussed its applicability to scalar quantum 
field theories in the introduction; in \cite{Fischer:2006vf} we discussed 
scaling in the gauge-Higgs theory. These applications may also be extended
to Yang-Mills theories in other gauges. \\[1ex]

{\bf Acknowledgments}\\
We are grateful for many discussions with Reinhard Alkofer, Markus Huber,
Felipe Llanes-Estrada, Kai Schwenzer and Lorenz von Smekal.
C.~F. was supported by the Helmholtz-University Young Investigator
Grant number VH-NG-332
and J.~M.~P. by Helmholtz Alliance HA216/EMMI.\\[1mm]

\begin{appendix}

\section{Integrated flow equation}\label{app:fun&scale}
\begin{figure}[h]
\centerline{\includegraphics[width=0.6\columnwidth]{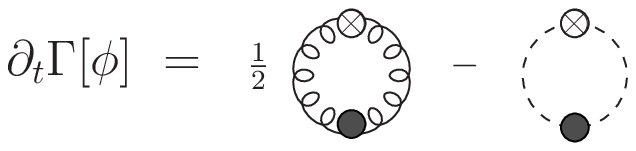}}
\caption{Functional renormalisation group equation (FRG) for the effective 
action. Filled circles denote fully dressed field dependent propagators. 
Crosses denote the regulator insertion $\partial_t R$. }
\label{fig:standardFRG}
\end{figure}

The standard form in the flow equation is depicted in 
Fig~\ref{fig:standardFRG}, 
where $t=\ln k$ is the logarithmic infrared scale and the cross
denotes the regulator insertion $\partial_t R$. If $R_k$ is chosen as
a mode cut-off and simply removes one momentum mode from the theory,
the loop integrations in the FRG equations for vertices is reduced to
a single loop momentum $q^2\simeq k^2$ and we can explicitly apply the
infrared power counting. This has been done in \cite{Fischer:2006vf}.
For the sake of the comparison with the DSE is also convenient to
rewrite the FRG as follows \cite{Pawlowski:2005xe},
\begin{widetext}
\begin{eqnarray}
  \partial_t \Gamma[\phi]=
  \012\Tr\left( \0{1}{\Gamma^{(2)}
      [\phi]+R}\partial_t R
  \right)=
  \012  \Tr \partial_t \ln 
  \left(\Gamma^{(2)}[\phi]+R\right)
  -\012\Tr\left( \0{1}{\Gamma^{(2)}[\phi]+R}\partial_t \Gamma^{(2)}\right)
  \,. 
\label{eq:rewrite} \end{eqnarray}

Upon integration from an initial momentum scale $k=\Lambda$ to $k=0$
this yields
\begin{eqnarray} 
  \Gamma[\phi]=
S_\Lambda[\phi]+
  \left(\012  \Tr \ln \Gamma^{(2)}[\phi]+{\rm ren}\right)%
-\012 \int_\Lambda^0 \0{dk'}{k'} \Tr \left(\0{1}{\Gamma^{(2)}[\phi]+R'} 
  \partial_{t'} \Gamma^{(2)} \right)+{\rm ren}
  \,,  
\label{eq:integrate}\end{eqnarray}
\end{widetext}
where $S_\Lambda=(\Gamma_\Lambda-\Tr \ln
(\Gamma^{(2)}[\phi]+R)_\Lambda -{\rm ren})$ entails the initial
condition at $k=\Lambda$ and the integral term on the rhs of
\eq{eq:integrate} is an RG-improvement term.  If we perform a momentum
rescaling as in \eq{eq:scalings} including that of $\Lambda$:
$\Lambda\to\lambda\Lambda$, the last term shows at most the same
scaling as the first one.  This entails up to renormalisation, that
\begin{equation}\label{eq:limlambda} 
  \lim_{\lambda\to 0}\Gamma[\phi] \simeq  \012  
  \Tr \ln \Gamma^{(2)}[\phi]\,, 
\end{equation}
as far as infrared scaling is concerned. Taking a gluon or ghost
derivative of \eq{eq:limlambda} leads to the diagrammatical
representation of the infrared asymptotics of the flow in
Fig.~\ref{fig:funFRG}.

\section{Vertices with two or more external scaling legs
  \label{app:vertices}}
  
In this appendix we investigate the kinematical scaling for vertices
$\Gamma^{2n,m}$ with $s$ soft gluon lines, $r_1$ soft ghost lines and
$r_2$ soft anti-ghost lines with $r=r_1+r_2$. Due to ghost-anti-ghost
symmetry we can always choose $r_2 \ge r_1$.  We will explore some
general situations and discuss exceptions at the end of this section.

First we discuss diagrams with isolated external legs with vanishing
momenta, i.e. diagrams where these legs are not neighbouring. These
are simple. In the case of isolated external ghost lines no
divergences are encountered due to \Eq{eq:10ghost}. Diagrams with
isolated external gluon lines scale like the corresponding diagrams
with one vanishing momentum, i.e. \Eq{eq:01ghost} if no hard external
gluon lines are present and \Eq{eq:01glue} otherwise. This can be
easily verified by going through some explicit examples. In case some
of the external gluonic momenta are parallel we always find
\Eq{eq:01glue}; the argument is similar to that given below
\Eq{eq:01glue}. Having said this, we concentrate on vertices with only
neighbouring legs with vanishing momenta for the remainder of this
section.

\begin{figure}[h]
\begin{center}
  \includegraphics[width=\columnwidth]{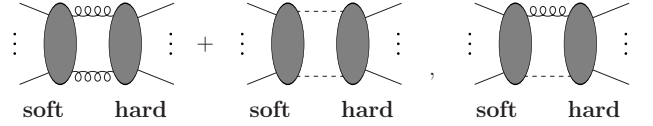}
  \caption{Diagram classes with kinematical scaling of only part of
    the vertices.}
 \label{fig:3class}
\end{center}
\end{figure}

In the following argument we concentrate on the FRG diagrams which are
only one loop. A part of the related diagrams decay into three classes
depicted in Fig.~\ref{fig:3class}. The other diagrams involve vertices
with soft and hard parts that cannot be separated in the above way. We
will evaluate these diagrams at the end of our discussion. The first
two classes of diagrams in Fig.~\ref{fig:3class} summarise possible
diagrams of neighbouring vanishing legs for $r_2 = r_1$ or
$r_2=r_1+2$. The third class summarises those with $r_2=r_1+1$.
The total scaling of a vertex with $r_1$ ghosts and
$r_2$ anti-ghosts with $r=r_1+r_2$ and $s$ gluons is given by
\begin{equation} \label{eq:totalrs}
t_{r,s}=d_{r+s}+\012 \left(r\kappa_{2,0}+s \kappa_{0,2}\right)\,,
\end{equation} 
where we have extended the scaling for ghost gluon vertices with
$r_1=r_2$ to non-existing vertices with $r_1\neq r_2$. We are now in
the position to deduce the scaling of vertices $\Gamma^{2n,m}$ with
$s+r_1+r_2$ vanishing external momenta for gluons+ghosts+anti-ghosts
respectively, by relating it to the global scaling \eq{eq:totalrs} of
a vertex with only vanishing external momenta: we simply remove the
hard parts of the diagrams and close the remaining open lines with
some combinations of soft vertices and propagators, the scaling of
which we then subtract.  What is left is the scaling $t_{2n,m}^{r,s}$
of the diagrams in Fig.~\ref{fig:3class}. Since this procedure is
insensitive to the specific combination of vertices added, we simply
take a minimal number of additional soft vertices for the explicit
computations, without loss of generality.

In the first diagram in Fig.~\ref{fig:3class} we substitute the hard 
part of the diagram by a soft full three gluon vertex, hence adding 
a further vanishing gluon momentum. From the resulting global scaling
of $t_{r,s+1}$ we have to subtract the global scaling $t_{0,3}$ of the
three-gluon vertex leading to 
\begin{equation}\label{eq:scaling1}
t^{r,s}_{2n,m}\leq t_{r,s+1}-t_{0,3}=t_{r,s}+\frac{2-d}{2}+2\kappa\,.
\end{equation} 
In \eq{eq:scaling1} we have used that $d_{r+s+1}-d_3= d_{r+s}-1$. As
already pointed out above, e.g.\ substituting the hard part of the diagram
with two ghost gluon vertices and one connecting ghost propagator
leads to the same result. The same is true for any other combination
of vertices in the hard part of the diagram. The $\leq$ in
\Eq{eq:scaling1} expresses the fact that there are cases where the
kinematical situation of the same vertex may be represented by the
first or second digram of Fig.~\ref{fig:3class} and either one may
carry the leading kinematical singularity.

To analyse the second diagram we have to distinguish two cases. For 
$r_1=r_2$ we substitute the hard part by a soft full ghost gluon vertex,
hence adding a further vanishing ghost momentum. Apart from
subtracting its global scaling $t_{2,1}=1/2$, we also have to take
care of the kinematical intricacies of ghost lines.  The hard part of
the second class of diagrams in Fig.~\ref{fig:3class} also carries a
momentum dependence that is linear in the momenta of the soft ghost
legs. This has to be added separately, and we arrive at
\begin{equation}\label{eq:scaling2}
t^{r,s}_{2n,m}\leq t_{r,s+1}+1-t_{2,1}=t_{r,s}-\kappa\,, 
\end{equation} 
where we have used that $d_{r+s+1}=d_{r+s}-(d-2)/4$. If $r_1=r_2-2$ the
two connecting lines between the soft and the hard part of the diagram
are two anti-ghosts and we have at least to substitute the hard part of
the diagram with two ghost gluon vertices and one connecting gluon
propagator. In our counting we again have to add two linear powers of
momenta due to the two connecting ghost legs. The net result is the
same as for the case $r_1=r_2$. Therefore the result \Eq{eq:scaling2}
with $r=r_1+r_2$ summarises both cases.

In the third diagram we substitute the hard part of the
diagram by a soft full ghost gluon vertex, hence adding a further
vanishing ghost momentum. Apart from subtracting its global scaling
$t_{2,1}=1/2$, the hard part of the third class of diagrams in
Fig.~\ref{fig:3class} also carries a momentum dependence that is
linear in the momentum of the soft ghost leg which has to be added
separately. We are led to 
\begin{equation}\label{eq:scaling3}
t^{r,s}_{2n,m}= t_{r+1,s}+\012-t_{2,1}=t_{r,s}-\0{d-2}{4}
+\012 \kappa\,.
\end{equation}
Note that $t^{r,s}_{2n,m}= t_{r+1,s}$ with $r+1=2 n$. 

We summarise the results of this section with the scaling relation 
\begin{equation}\label{eq:scalingsumB}
t^{r,s}_{2n,m}=
t_{r,s}+\left\{\begin{array}{lc} 
-\0{d-2}{4}
+\012 \kappa \qquad 
& r_2= r_1+1\,,\\[2ex]
\min\left(-\kappa, \frac{2-d}{2}+2\kappa \right) \qquad 
& {\rm else} \end{array}\right. 
\end{equation}
where we have used \eq{eq:scaling1},\eq{eq:scaling2} and
\eq{eq:scaling3}. Note that one can swap between the different classes
of diagrams by isolating one or two soft ghost or anti-ghost vertices
within the hard part of the diagrams. This effectively removes these
vertices from the counting. However, since the scaling relations is
different for the three classes it could pay off in a more singular
behaviour. It is easy to check that this is not the case, and hence
\eq{eq:scalingsumB} represents the full maximal scaling of the diagram 
classes depicted in Fig.~\ref{fig:3class}.  

\begin{figure}[h]
\begin{center}
  \includegraphics[width=0.4\columnwidth]{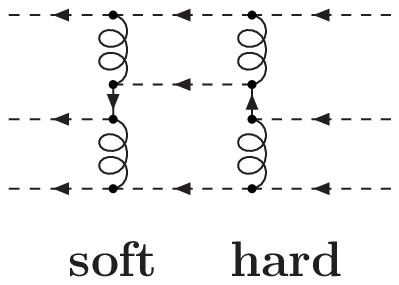}
  \caption{Example from an exceptional class of diagrams 
  with kinematical scaling of only part of the vertices.}
 \label{fig:irreducible}
\end{center}
\end{figure}

We close with a discussion of the remaining diagram classes. A
interesting specific case is depicted in Fig.~\ref{fig:irreducible}.
This seems to entail that one can have neighbouring ghost vertices or
anti-ghost vertices. However, one can show that this only is possible
at the expense of additional loops as is present in
Fig.~\ref{fig:irreducible}. At its core this relates to the fact that
the ghost number of vertices $\Gamma^{2n,m}$ vanishes. The scaling of 
the related sub-diagram is always positive, in the present case it is 
$1/2$. Restricting ourselves again to the case $r_1\leq r_2$, we 
deduce that for $r_2>r_1+2$ we are effectively reduced to the soft scaling 
of a diagram with $r_1$ ghost and $r_1+2$ anti-ghost legs and hence
\Eq{eq:scalingsumB} applies. Because of 
ghost--anti-ghost symmetry this covers the general case.

\end{appendix}


\begin{thebibliography}{99}

\bibitem{Kugo}
T.\ Kugo and I.\ Ojima,
Prog.\ Theor.\ Phys.\ Suppl.\ {\bf 66}, 1 (1979) [Erratum Prog.\ Theor.\ Phys.\ {\bf 71}, 1121 (1984)];
T.\ Kugo,
arXiv:hep-th/9511033.

\bibitem{Gribov}
  V.\ N.\ Gribov,
  Nucl.\ Phys.\ B {\bf 139}, 1 (1978).

\bibitem{gzwanziger2}
  D.\ Zwanziger,
  Phys.\ Lett.\ B {\bf 257}, 168 (1991);
  Nucl.\ Phys.\ B {\bf 364}, 127 (1991);
  Nucl.\ Phys.\ B {\bf 412}, 657 (1994);
  Phys.\ Rev.\ D {\bf 65}, 094039 (2002) [arXiv: hep-th/0109224].

\bibitem{von Smekal:1997vx}
  L.~von Smekal, A.~Hauck and R.~Alkofer,
  Annals Phys.\  {\bf 267} (1998) 1
  [Erratum-ibid.\  {\bf 269} (1998) 182]
  [arXiv:hep-ph/9707327];
  L.~von Smekal, R.~Alkofer and A.~Hauck,
  Phys.\ Rev.\ Lett.\  {\bf 79} (1997) 3591
  [arXiv:hep-ph/9705242].

\bibitem{Lerche:2002ep}
  C.~Lerche and L.~von Smekal,
  Phys.\ Rev.\  D {\bf 65}, 125006 (2002)
  [arXiv:hep-ph/0202194].

\bibitem{Pawlowski:2003hq}
 J.~M. Pawlowski, D.~F. Litim, S.~Nedelko, and L.~von Smekal {\em Phys. Rev.
   Lett.} {\bf 93} (2004) 152002 [hep-th/0312324];
   AIP Conf.\ Proc.\  {\bf 756} (2005) 278
   [hep-th/0412326].

\bibitem{Alkofer:2004it}
  R.\ Alkofer, C.\ S.\ Fischer and F.\ J.\ Llanes-Estrada,
  Phys.\ Lett.\ B {\bf 611}, 279 (2005)
  [arXiv:hep-th/0412330].

\bibitem{Fischer:2006vf}
  C.~S.~Fischer and J.~M.~Pawlowski,
  Phys.\ Rev.\  D {\bf 75} (2007) 025012
  [arXiv:hep-th/0609009].

\bibitem{Huber:2007kc}
  M.~Q.~Huber, R.~Alkofer, C.~S.~Fischer and K.~Schwenzer,
  Phys.\ Lett.\  B {\bf 659} (2008) 434
  [arXiv:0705.3809 [hep-ph]].

\bibitem{Fischer:2008uz}
  C.~S.~Fischer, A.~Maas and J.~M.~Pawlowski,
  arXiv:0810.1987 [hep-ph].

\bibitem{Cornwall:1981zr}
  J.~M.~Cornwall,
  Phys.\ Rev.\  D {\bf 26} (1982) 1453.


\bibitem{Aguilar:2008xm}  
  A.~C.~Aguilar, D.~Binosi and J.~Papavassiliou,
  Phys.\ Rev.\  D {\bf 78}, 025010 (2008)
  [arXiv:0802.1870 [hep-ph]].

\bibitem{Boucaud:2008ji}
  Ph.~Boucaud, J.~P.~Leroy, A.~L.~Yaouanc, J.~Micheli, O.~Pene and J.~Rodriguez-Quintero,
  JHEP {\bf 0806}, 012 (2008)
  [arXiv:0801.2721 [hep-ph]].

\bibitem{Dudal:2008sp}
  D.~Dudal, J.~A.~Gracey, S.~P.~Sorella, N.~Vandersickel and H.~Verschelde,
  arXiv:0806.4348 [hep-th].

\bibitem{Alkofer:2008jy}
  R.~Alkofer, M.~Q.~Huber and K.~Schwenzer,
  arXiv:0801.2762 [hep-th];
  arXiv:0812.4045 [hep-ph].


\bibitem{Pawlowski:2005xe}
  J.~M.~Pawlowski,
  Annals Phys.\  {\bf 322} (2007) 2831
  [arXiv:hep-th/0512261].

\bibitem{Pawlowski:2001df}
  J.~M.~Pawlowski,
  Int.\ J.\ Mod.\ Phys.\  A {\bf 16} (2001) 2105.

\bibitem{Zwanziger:2001kw}
  D.~Zwanziger,
  Phys.\ Rev.\  D {\bf 65} (2002) 094039
  [arXiv:hep-th/0109224].

\bibitem{Fischer:2002hna}
  C.~S.~Fischer and R.~Alkofer,
  Phys.\ Lett.\  B {\bf 536} (2002) 177
  [arXiv:hep-ph/0202202].

\bibitem{Maas:2004se}
  A.~Maas, J.~Wambach, B.~Gruter and R.~Alkofer,
  Eur.\ Phys.\ J.\  C {\bf 37} (2004) 335
  [arXiv:hep-ph/0408074].

\bibitem{Ilgenfritz:2006he}
  E.~M.~Ilgenfritz, M.~Muller-Preussker, A.~Sternbeck, A.~Schiller and I.~L.~Bogolubsky,
  Braz.\ J.\ Phys.\  {\bf 37}, 193 (2007)
  [arXiv:hep-lat/0609043].

\bibitem{Cucchieri:2008qm}
  A.~Cucchieri, A.~Maas and T.~Mendes,
  Phys.\ Rev.\  D {\bf 77}, 094510 (2008)
  [arXiv:0803.1798 [hep-lat]].

\bibitem{Schleifenbaum:2004id}
  W.\ Schleifenbaum, A.\ Maas, J.\ Wambach and R.\ Alkofer,
  Phys.\ Rev.\ D {\bf 72}, 014017 (2005)
  [arXiv:hep-ph/0411052].

\bibitem{Fischer:2004uk}
  C.~S.~Fischer and H.~Gies,
  JHEP {\bf 0410}, 048 (2004)
  [arXiv:hep-ph/0408089].

\bibitem{Kellermann:2008iw}
  C.~Kellermann and C.~S.~Fischer,
  Phys.\ Rev.\  D {\bf 78}, 025015 (2008)
  [arXiv:0801.2697 [hep-ph]].

\bibitem{Bloch:2003yu}
  J.~C.~R.~Bloch,
  Few Body Syst.\  {\bf 33}, 111 (2003)
  [arXiv:hep-ph/0303125].


\end{thebibliography}
 \end{document}